\documentstyle[twoside,fleqn,espcrc2,epsfig]{article}

\def\rmd{{\rm d}}

\def\rme{{\rm e}}
\def\rmO{{\rm O}}

\def\defeq{\mathrel{\mathop=^{\rm def}}}
\def\proof{\noindent{\sl Proof:}\kern0.6em}

\def\frac#1#2{\hbox{$#1\over#2$}}
\def\dual{\mathstrut^*\kern-0.1em}

\def\lvec#1{\setbox0=\hbox{$#1$}
    \setbox1=\hbox{$\scriptstyle\leftarrow$}
    #1\kern-\wd0\smash{
    \raise\ht0\hbox{$\raise1pt\hbox{$\scriptstyle\leftarrow$}$}}
    \kern-\wd1\kern\wd0}
\def\rvec#1{\setbox0=\hbox{$#1$}
    \setbox1=\hbox{$\scriptstyle\rightarrow$}
    #1\kern-\wd0\smash{
    \raise\ht0\hbox{$\raise1pt\hbox{$\scriptstyle\rightarrow$}$}}
    \kern-\wd1\kern\wd0}

\def\nabstar#1{\nabla\kern-0.5pt\smash{\raise 4.5pt\hbox{$\ast$}}
               \kern-4.5pt_{#1}}

\def\drvstar#1{\partial\kern-0.5pt\smash{\raise 4.5pt\hbox{$\ast$}}
               \kern-5.0pt_{#1}}

\def\GeV{{\rm GeV}}

\def\Nf{N_{\rm f}}
\def\psibar{\overline{\psi}}

\def\rhoprime{\rho\kern1pt'}
\def\rhobar{\bar{\rho}}
\def\rhobarprime{\rhobar\kern1pt'}
\def\rhobartilde{\kern2pt\tilde{\kern-2pt\rhobar}}
\def\rhobartildeprime{\kern2pt\tilde{\kern-2pt\rhobar}\kern1pt'}

\def\zetabar{\bar{\zeta}}
\def\zetaprime{\zeta\kern1pt'}
\def\zetabarprime{\zetabar\kern1pt'}
\def\zetar{\zeta_{\raise-1pt\hbox{\sixrm R}}}
\def\zetabarr{\zetabar_{\raise-1pt\hbox{\sixrm R}}}

\def\phiimpr{\phi_{\kern0.5pt\hbox{\sixrm I}}}

\def\diracstar#1#2{
    \setbox0=\hbox{$\gamma$}\setbox1=\hbox{$\gamma_{#1}$}
    \gamma_{#1}\kern-\wd1\kern\wd0
    \smash{\raise4.5pt\hbox{$\scriptstyle#2$}}}

\def\ba{b_{\rm A}}
\def\bv{b_{\rm V}}
\def\bp{b_{\rm P}}

\def\bg{b_{\rm g}}
\def\bm{b_{\rm m}}

\def\ca{c_{\rm A}}
\def\cv{c_{\rm V}}

\def\csw{c_{\rm sw}}

\def\fp{f_{\rm P}}

\def\f1{f_1}

\def\opprime#1{\setbox0=\hbox{${\cal O}$}\setbox1=\hbox{${\cal O}_{\rm #1}$}
    {\cal O}_{\rm #1}\kern-\wd1\kern\wd0
    \smash{\raise4.5pt\hbox{\kern1pt$\scriptstyle\prime$}}\kern1pt}

\def\ophatprime#1{\setbox0=\hbox{$\widehat{\cal O}$}
    \setbox1=\hbox{$\widehat{\cal O}_{\rm #1}$}
    \widehat{\cal O}_{\rm #1}\kern-\wd1\kern\wd0
    \smash{\raise4.5pt\hbox{\kern1pt$\scriptstyle\prime$}}\kern1pt}

\def\bopprime#1{\setbox0=\hbox{${\cal O}$}\setbox1=\hbox{${\cal O}_{\rm #1}$}
    {\cal L}_{\rm #1}\kern-\wd1\kern\wd0
    \smash{\raise4.5pt\hbox{\kern1pt$\scriptstyle\prime$}}\kern1pt}

\def\blagprime#1{\setbox0=\hbox{${\cal B}$}\setbox1=\hbox{${\cal B}_{#1}$}
    {\cal B}_{#1}\kern-\wd1\kern\wd0
    \smash{\raise5.2pt\hbox{\kern1pt$\scriptstyle\prime$}}\kern1pt}

\def\gbar{\bar{g}}

\def\mq{m_{\rm q}}
\def\mqtilde{\widetilde{m}_{\rm q}}

\def\mc{m_{\rm c}}

\def\za{Z_{\rm A}}
\def\zp{Z_{\rm P}}
\def\zv{Z_{\rm V}}
\def\zg{Z_{\rm g}}
\def\zm{Z_{\rm m}}

\def\gtilde{\tilde{g}_0}

\def\MSbar{\overline{\rm MS}}

\def\msbar{{\rm \overline{MS\kern-0.05em}\kern0.05em}}

\def\rmd{{\rm d}}

\def\rme{{\rm e}}
\def\rmO{{\rm O}}

\def\bfx{{\bf x}}
\def\bfy{{\bf y}}
\def\bfz{{\bf z}}

\def\defeq{\mathrel{\mathop=^{\rm def}}}
\def\proof{\noindent{\sl Proof:}\kern0.6em}

\def\frac#1#2{\hbox{$#1\over#2$}}
\def\dual{\mathstrut^*\kern-0.1em}

\def\lvec#1{\setbox0=\hbox{$#1$}
    \setbox1=\hbox{$\scriptstyle\leftarrow$}
    #1\kern-\wd0\smash{
    \raise\ht0\hbox{$\raise1pt\hbox{$\scriptstyle\leftarrow$}$}}
    \kern-\wd1\kern\wd0}
\def\rvec#1{\setbox0=\hbox{$#1$}
    \setbox1=\hbox{$\scriptstyle\rightarrow$}
    #1\kern-\wd0\smash{
    \raise\ht0\hbox{$\raise1pt\hbox{$\scriptstyle\rightarrow$}$}}
    \kern-\wd1\kern\wd0}

\def\nabstar#1{\nabla\kern-0.5pt\smash{\raise 4.5pt\hbox{$\ast$}}
               \kern-4.5pt_{#1}}

\def\drvstar#1{\partial\kern-0.5pt\smash{\raise 4.5pt\hbox{$\ast$}}
               \kern-5.0pt_{#1}}

\def\momp#1#2{
    \setbox0=\hbox{${#1}$}\setbox1=\hbox{${#1}_{#2}$}
    {#1}_{#2}\kern-\wd1\kern\wd0
    \smash{\raise4.5pt\hbox{$\scriptscriptstyle +$}}}
\def\momm#1#2{
    \setbox0=\hbox{${#1}$}\setbox1=\hbox{${#1}_{#2}$}
    {#1}_{#2}\kern-\wd1\kern\wd0
    \smash{\raise4.5pt\hbox{$\scriptscriptstyle -$}}}
\def\mompm#1#2{
    \setbox0=\hbox{${#1}$}\setbox1=\hbox{${#1}_{#2}$}
    {#1}_{#2}\kern-\wd1\kern\wd0
    \smash{\raise4.5pt\hbox{$\scriptscriptstyle \pm$}}}
\def\smomp#1#2{
    \setbox0=\hbox{${#1}$}\setbox1=\hbox{${#1}_{#2}$}
    {#1}_{#2}\kern-\wd1\kern\wd0
    \smash{\raise3pt\hbox{$\scriptscriptstyle +$}}}
\def\smomm#1#2{
    \setbox0=\hbox{${#1}$}\setbox1=\hbox{${#1}_{#2}$}
    {#1}_{#2}\kern-\wd1\kern\wd0
    \smash{\raise3pt\hbox{$\scriptscriptstyle -$}}}
\def\smompm#1#2{
    \setbox0=\hbox{${#1}$}\setbox1=\hbox{${#1}_{#2}$}
    {#1}_{#2}\kern-\wd1\kern\wd0
    \smash{\raise3pt\hbox{$\scriptscriptstyle \pm$}}}

\def\si{\kern1pt{\rm si}}
\def\co{\kern1pt{\rm co}}

\def\GeV{{\rm GeV}}

\def\Nf{N_{\rm f}}
\def\psibar{\overline{\psi}}

\def\rhoprime{\rho\kern1pt'}
\def\rhobar{\bar{\rho}}
\def\rhobarprime{\rhobar\kern1pt'}
\def\rhobartilde{\kern2pt\tilde{\kern-2pt\rhobar}}
\def\rhobartildeprime{\kern2pt\tilde{\kern-2pt\rhobar}\kern1pt'}

\def\zetabar{\bar{\zeta}}
\def\zetaprime{\zeta\kern1pt'}
\def\zetabarprime{\zetabar\kern1pt'}
\def\zetar{\zeta_{\raise-1pt\hbox{\sixrm R}}}
\def\zetabarr{\zetabar_{\raise-1pt\hbox{\sixrm R}}}

\def\phiimpr{\phi_{\kern0.5pt\hbox{\sixrm I}}}

\def\diracstar#1#2{
    \setbox0=\hbox{$\gamma$}\setbox1=\hbox{$\gamma_{#1}$}
    \gamma_{#1}\kern-\wd1\kern\wd0
    \smash{\raise4.5pt\hbox{$\scriptstyle#2$}}}

\def\ba{b_{\rm A}}

\def\bp{b_{\rm P}}

\def\bv{b_{\rm V}}

\def\bg{b_{\rm g}}
\def\bm{b_{\rm m}}

\def\ca{c_{\rm A}}
\def\cv{c_{\rm V}}

\def\csw{c_{\rm sw}}

\def\fp{f_{\rm P}}

\def\f1{f_1}

\def\h1{h_1}

\def\opprime#1{\setbox0=\hbox{${\cal O}$}\setbox1=\hbox{${\cal O}_{\rm #1}$}
    {\cal O}_{\rm #1}\kern-\wd1\kern\wd0
    \smash{\raise4.5pt\hbox{\kern1pt$\scriptstyle\prime$}}\kern1pt}

\def\ophatprime#1{\setbox0=\hbox{$\widehat{\cal O}$}
    \setbox1=\hbox{$\widehat{\cal O}_{\rm #1}$}
    \widehat{\cal O}_{\rm #1}\kern-\wd1\kern\wd0
    \smash{\raise4.5pt\hbox{\kern1pt$\scriptstyle\prime$}}\kern1pt}

\def\bopprime#1{\setbox0=\hbox{${\cal O}$}\setbox1=\hbox{${\cal O}_{\rm #1}$}
    {\cal L}_{\rm #1}\kern-\wd1\kern\wd0
    \smash{\raise4.5pt\hbox{\kern1pt$\scriptstyle\prime$}}\kern1pt}

\def\blagprime#1{\setbox0=\hbox{${\cal B}$}\setbox1=\hbox{${\cal B}_{#1}$}
    {\cal B}_{#1}\kern-\wd1\kern\wd0
    \smash{\raise5.2pt\hbox{\kern1pt$\scriptstyle\prime$}}\kern1pt}

\def\gbar{\bar{g}}

\def\muq{\mu_{\rm q}}
\def\mq{m_{\rm q}}
\def\mqtilde{\widetilde{m}_{\rm q}}

\def\mc{m_{\rm c}}

\def\za{Z_{\rm A}}
\def\zv{Z_{\rm V}}
\def\zp{Z_{\rm P}}

\def\zg{Z_{\rm g}}
\def\zm{Z_{\rm m}}

\def\Za{\za}
\def\Zv{\zv}
\def\Zp{\zp}

\def\Zg{\zg}
\def\Zm{\zm}

\def\gtilde{\tilde{g}_0}

\def\msbar{{\rm \overline{MS\kern-0.05em}\kern0.05em}}

\title{Non-perturbative Renormalization 
in Lattice Field Theory\thanks{based on a plenary talk 
presented at the International Symposium on 
Lattice Field Theory, August 17 -- 22, 2000, Bangalore, India.
$^\dagger$address after October 1, 2000: 
CERN, Theory Division, CH-1211 Geneva 23, Switzerland}}

\author{Stefan Sint$^\dagger$ \address{Universit\`a di Roma ``Tor
Vergata'', Dipartimento di Fisica, Via della Ricerca Scientifica
                      1, I-00133 Roma, Italy}
}

\begin{document}
\begin{abstract}

I review the strategies which have been developped in recent years 
to solve the non-perturbative renormalization 
problem in lattice field theories. Although the techniques are
general, the  focus will be on applications to lattice QCD.
I discuss the momentum subtraction and finite volume schemes,
and their application to scale dependent renormalizations.
The problem of finite renormalizations is illustrated 
with the example of explicit chiral symmetry breaking,
and I give a short status report concerning Symanzik's
improvement programme to O($a$).

\end{abstract}
\maketitle

\section{INTRODUCTION}

A Quantum Field Theory is renormalized by
defining the infinite cutoff or continuum limit
of the regularized theory. The procedure is 
familiar in perturbation theory, but is more
generally applicable. In particular, the renormalization 
of asymptotically free theories at low 
scales requires a genuinely non-perturbative approach.
Unfortunately there exist very few analytical tools,
and numerical simulations of the lattice regularized theory
are often the only method to obtain quantitative results.

One is thus led to discuss renormalization of the lattice
field theory and to take the continuum limit based
on numerical data. In general this is only possible
by making assumptions about the continuum approach.
Motivated by Symanzik's analysis of the cutoff dependence in
perturbation theory, one usually assumes 
that the contiuum limit is reached with power corrections 
in the lattice spacing $a$, possibly modified by logarithms. 
In this context ``unexpected results'' have been
presented recently by Hasenfratz and 
Niedermayer~\cite{Hase_Nieder,Niedermayer}.
They find that the continuum approach of a renormalized coupling
in the O(3) non-linear sigma model is compatible 
with being linear in $a$, rather than quadratic, 
as one would expect for a bosonic model.  While 
the statistical precision of the data is quite impressive,
it seems premature to draw conclusions. We also note
that there are many results in (quenched) QCD which are
well compatible with expectations, and some of them will be
presented below.  Finally, I will  assume the ```standard wisdom'' 
concerning asymptotic freedom and the operator product 
expansion (OPE) to be correct. While this has never 
been established beyond perturbation theory, a numerical
check of the OPE in the O(3) model has recently been presented 
in ref.~\cite{Caracciolo_et_al}. For an unconventional
point of view regarding asymptotic freedom I refer to 
ref.~\cite{Seiler_lat00}.

Despite the more general title I will 
focus on lattice QCD. For one,  
non-perturbative renormalization techniques in QCD
have reached a mature stage, so that systematic errors
can be discussed. Second, QCD is a typical example,
and the techniques carry over essentially unchanged to
other theories. 

This article is organized as follows. In sect.~2,  
I sketch the non-perturbative renormalization procedure for QCD. 
Then the two classes of schemes are reviewed which have been
proposed to solve the problem of scale-dependent renormalization, 
namely the momentum subtraction schemes (RI/MOM, sect.~3)  
and finite volume schemes derived from the Schr\"odinger functional (sect.~4).
This is followed by a discussion of finite renormalizations (sect.~5), 
and  O($a$) improvement (sect. 6). I conclude with a 
remark concerning power divergences.

\section{NON-PERTURBATIVE RENORMALIZATION OF QCD}

\subsection{Determination of fundamental parameters}

The free parameters of the QCD action are the bare coupling 
and the bare quark mass parameters. 
At low energies the theory is usually renormalized in a 
hadronic scheme, i.e.~one chooses a corresponding number of
hadronic observables which are kept fixed as the continuum
limit is taken.

Once the theory has been renormalized in this way, no freedom is left and
the renormalized parameters in any other scheme are predicted
by the theory. Of particular interest is the relation to
the renormalized coupling and quark masses 
in a perturbative scheme. This is equivalent to a determination
of the $\Lambda$-parameter and the renormalization 
group invariant quark masses. For, given the 
running coupling $\bar{g}$ and quark masses $\overline{m}_i$
at scale $\mu$, one finds the {\em exact} relations
\begin{eqnarray}
  \Lambda &=& \mu\,(b_0\gbar^2)^{-b_1/2b_0^2}\,\exp\left\{
   -\frac{1}{2b_0\gbar^2}\right\}
    \label{Lambda} \\
   &&\times\exp\left\{-\int_0^{\gbar}
   \rmd x \left[\frac{1}{\beta(x)}
              +\frac{1}{b_0x^3}
              -\frac{b_1}{b_0^2x}\right]\right\}, 
   \nonumber\\[1ex]
  M_i &=& \overline{m}_i\,(2b_0\gbar^2)^{-d_0/2b_0}
 \nonumber\\
  &&\times \exp\left\{-\int_0^{\gbar} 
  \rmd x\left[\frac{\tau(x)}{\beta(x)}-\frac{d_0}{b_0x}\right]\right\},
  \label{RGImass}
\end{eqnarray}
where $i$ labels the quark flavours. Here, $\beta$ and $\tau$ are
renormalization group functions with asymptotic expansions,
\begin{eqnarray}
  \beta(g)&\sim& -b_0g^3-b_1g^5+\ldots,
  \label{beta}\\
  \tau(g)&\sim& -d_0g^2-d_1g^4+\ldots.
  \label{tau}
\end{eqnarray}
Note that the renormalization group invariant quark masses are 
scheme independent, whereas the Lambda parameter changes
according to
\begin{equation}
  \Lambda' = \Lambda\exp\{c^{(1)}/b_0\}.
\end{equation}
Here, $c^{(1)}$ is the one-loop coefficient relating the 
renormalized coupling constants $g'=g+c^{(1)}g^3 +\rmO(g^5)$.
$\Lambda$ in the $\MSbar$ scheme and $M_i$
are referred to as the fundamental parameters of QCD.

In order to relate the hadronic scheme to these parameters 
one may introduce an intermediate renormalization 
scheme involving quantities which can be evaluated both in perturbation 
theory and non-perturbatively. An obvious possibility is 
to take a renormalized coupling and renormalized 
quark masses which are defined beyond perturbation 
theory. In such a scheme also the renormalization group
functions $\beta$ and $\tau$ are defined non-perturbatively,
so that the formulae (\ref{Lambda},\ref{RGImass}) 
can be applied  at any scale $\mu$, once the 
matching to the hadronic scheme has been performed.

\subsection{Renormalization of composite operators}

In the Standard Model composite operators 
often arise when the OPE is used to 
separate widely different scales in a physical amplitude. 
While the operators are originally defined
at high energies (e.g. at $\mu=M_W$) and in a perturbative
scheme,  one is usually interested in their matrix
elements between hadronic states.
This defines a matching problem between a perturbative
renormalization scheme for the operator (e.g. the $\MSbar$
scheme) and a hadronic scheme, where a particular hadronic
matrix element of the operator assumes a prescribed value.

The problem may again be approached by defining 
an intermediate renormalization scheme which can be evaluated
both non-perturbatively and in perturbation theory.
At high energies one may then either convert to a perturbative
scheme, or one may construct the renormalization
group invariant operator, in close analogy to the fundamental
parameters introduced above.

\subsection{The problem of scales}

The main problem for the lattice approach is due to the large
scale differences. In order to measure hadron
masses the hadrons have to fit into the (necessarily finite)
space-time volume, i.e.~$ L^{-1} \ll m_\pi$.
On the other hand, the matching to the intermediate scheme is done
via the bare parameters at a matching scale $\mu_0$, and one 
requires $ \mu_0 \ll a_{\rm max}^{-1}$,
where $a_{\rm max}$ denotes the largest lattice spacing considered in the 
continuum extrapolation of the matching condition. 
As the available lattice sizes $L/a$ are limited, $\mu_0$ can   
not be too large, and perturbation theory at this scale may not be reliable.
This ought to be checked by tracing the non-perturbative scale evolution of 
the renormalized parameters in the intermediate scheme. 
If the on-set of perturbative evolution is observed around a scale $\mu$,
one may then apply the formulae (\ref{Lambda},\ref{RGImass}) using 
the perturbative expressions~(\ref{beta},\ref{tau}).

Combining the various requirements, one gets
\begin{equation}
   L^{-1} \ll m_\pi, \Lambda \ll \mu \ll a^{-1},
  \label{scales}
\end{equation}
and one also wants to vary $a$ in order to take the continuum limit.
A naive estimate then leads to ratios $L/a =\rmO(10^3)$  which are 
clearly too large to be realised on a single lattice.

\subsection{Shortcut methods}

Given a hadronic scheme at fixed $g_0$ one may be tempted to
avoid introducing the intermediate renormalization scheme
by considering the bare coupling and the bare quark masses
themselves as running parameters defined at the 
cutoff scale $a^{-1}$. One may then match
directly to, say, the $\MSbar$ scheme at scale $\mu=a^{-1}$,
by using bare perturbation theory~\cite{others}.
While this method is quick, it is also very difficult
to estimate the error in a reliable way. 
In particular it is impossible to disentangle renormalization
from cutoff effects, as any variation of $a$ is at the same
time a variation of the renormalization scale. 

\section{THE RI/MOM SCHEME}

The RI/MOM scheme~\cite{MOM} (RI for ``regularization independent'')
is a momentum subtraction scheme applied 
in a non-perturbative 
context~\cite{MOM,MOM1,MOM2,MOM3,MOM4,MOM5,MOM6,MOM7,MOM8}.
It plays the r\^ole of an intermediate renormalization
scheme in the sense of the previous section.

\subsection{The renormalization procedure}

In the RI/MOM scheme one considers correlation functions
with external quark states in momentum space.
A necessary first step therefore consists in fixing the gauge
(usually Landau gauge). One then follows the same
procedure as in perturbation theory: first the full quark 
propagator is computed for a set of momenta
and then used to amputate the external legs of the correlation function
of interest. The resulting vertex function is then
considered as a function of a single momentum,
and one imposes that its renormalized counterpart be equal
to its tree-level expression at the subtraction point, 
viz.
\begin{equation}
   \Gamma_{\rm R}(p)\bigl\vert_{p^2=\mu^2} =\Gamma(p)^{\rm tree}.
 \label{MOM_cond}
\end{equation}
For example, in the case of a multiplicatively renormalizable quark bilinear
operator $O$ eq.~(\ref{MOM_cond}) determines the operator renormalization
constant $Z_O(g_0,a\mu)$  up to the quark wave function renormalization.
The latter is then determined by considering the vertex function of 
a conserved current~\cite{MOM}.
Note that the renormalization constant is usually assumed to be 
quark mass independent. This can be achieved by 
imposing the renormalization condition
(\ref{MOM_cond}) in the chiral limit~\cite{Weinberg}.

The RI/MOM scheme may also be used to obtain non-perturbatively
defined  renormalized parameters. The PCAC and PCVC relations
imply that renormalized quark masses can be defined 
using the inverse renormalization constants
of the axial or the scalar density. To define 
a renormalized gauge coupling one may allow 
for external gluon states and impose a RI/MOM
condition on  the 3-gluon vertex function~\cite{Parrinello,Alles1}.

\subsection{Matching to other schemes}

Suppose one is interested in a hadronic matrix element 
of some operator $O$ defined in the $\MSbar$ scheme.
As a first step one relates the $\MSbar$ scheme 
to the RI/MOM scheme by inserting the renormalized operator
in the renormalization condition~(\ref{MOM_cond}).
This determines the finite renormalization constant $Z_O^{\MSbar}$
relating the operators in the $\MSbar$ and RI/MOM
schemes, order by order in the renormalized coupling $g^2_{\MSbar}(\mu)$.

Next, one computes the hadronic matrix element 
${\cal M}(a)=\langle h'| O |h\rangle$ for a sequence 
of lattice spacings and sets
\begin{equation}
  Z_O^{\MSbar}\left(g^2_{\MSbar}(\mu)\right) {\cal M}_{\MSbar}(\mu) 
  \defeq
  \lim_{a\rightarrow 0}  Z_O(a\mu){\cal M}(a).
\end{equation}
where $\mu$ is fixed in terms of a hadronic scale and
all mass parameters are supposed to be renormalized in a 
hadronic scheme.
Here one has used the regularization independence of the RI/MOM
condition to {\em define} the  l.h.s.~of this equation.
Provided one has previously determined
$g_{\MSbar}(\mu)$ this completely specifies the
matrix element in the $\MSbar$ scheme to the calculated
order in perturbation theory. Of course, it is assumed
that continuum perturbation theory is applicable at the scale $\mu$. 
Whether this is the case can be checked by repeating the
matching procedure at a different scale $\mu'$.
If the change is compatible with perturbative evolution from
$\mu$ to $\mu'$ one may use perturbation
theory to run to larger scales and extract the renormalization
group invariant matrix element. 

\begin{figure}[htb]
\epsfig{file=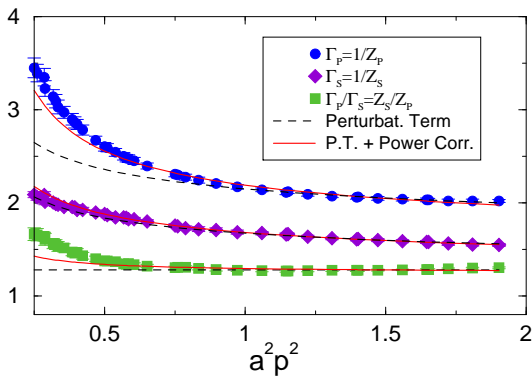,
       width=7cm}%
\vskip -5ex
\caption{Non-perturbative vs.~perturbative
scale dependence of the renormalization constants of 
the scalar and axial density (ref.~\cite{MOM6},
improved Wilson fermions at $\beta=6.2$).}
\vskip -4ex
\label{Rome}
\end{figure}

\begin{figure}[htb]
\epsfig{file=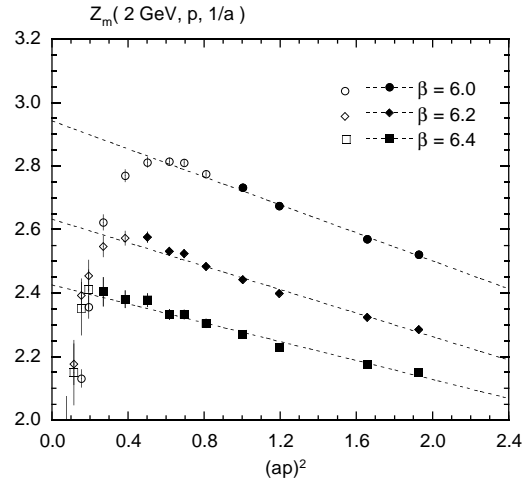,
       width=7cm}%
\vskip -5ex
\caption{Renormalization of the quark mass with Kogut-Susskind 
fermions~\cite{MOM5}. $Z_m$ should be $p$ independent 
if continuum perturbation theory was appropriate.}
\vskip -4ex
\label{Aoki_Zmq}
\end{figure}
\subsection{Examples}

Typical examples of operator renormalizations are given in
fig.~\ref{Rome}, where the renormalization constants
for the non-singlet axial and scalar densities are plotted
against the renormalization scale, for O($a$) improved
Wilson quarks in the quenched approximation at $\beta=6.2$~\cite{MOM6}. 
Also plotted are the curves obtained in {\em continuum}
perturbation theory. As one observes good agreement with the perturbative
curves, $p$-dependent cutoff effects are small and perturbation
theory seems to apply at  
scales around $\mu\simeq 1/a(\beta=6.2)$.

In order to determine the light quark masses,
the JLQCD collaboration renormalizes the
scalar density using (quenched) Kogut-Susskind fermions 
at 3 $\beta$-values corresponding  roughly to a variation
of the cutoff by a factor of 2~\cite{MOM5}.
The renormalization condition is extrapolated to the chiral
limit, and the renormalized quark mass is related 
to the bare mass $m$ through
\begin{equation}
  m_{\MSbar}(\mu_0)=\lim_{a\rightarrow 0}Z_m(\mu_0,p,1/a)\,m
 \label{mKS}
\end{equation}
The total renormalization factor $Z_m$ consists
of the non-perturbatively defined
renormalization constant at scale $p$ 
and the perturbative 3-loop evolution factor 
relating the scales $p$ and $\mu_0$. If {\em continuum}
perturbation theory was adequate between the
scales $\mu_0=2\,\GeV$ and $p$, then $Z_m$ should be independent
of $p$. This is not the case as fig.~\ref{Aoki_Zmq}
demonstrates. However, if the continuum 
extrapolation of eq.~(\ref{mKS}) is
performed for 2 different momenta $p=1.8\,\GeV$ and $p=2.6\,\GeV$,
the results are nicely compatible, demonstrating that
perturbation theory adequately describes the scale evolution
in this range of scales. In particular, 
the window condition $\Lambda\ll\mu$ seems to be satisfied
despite the absence of plateaux in fig.~\ref{Aoki_Zmq} at fixed $\beta$.

Further examples can be found
in~\cite{MOM1,MOM2,MOM3,MOM4,MOM5,MOM6,MOM7,MOM8}.
In particular, the RBC collaboration has applied
this scheme to the renormalization of operators 
in lattice QCD with Domain-Wall fermions~\cite{MOM7,MOM8}.

\subsection{Merits and problems of the RI/MOM scheme}

The RI/MOM scheme provides a versatile framework which
can be easily adapted to new renormalization problems.
Due to the regularization independence, 
perturbative calculations can be performed in a
continuum scheme, with many results being already 
available.  
Furthermore, the use of external quark states
is not only helpful in perturbation theory, but also leads
to good signals in numerical simulations.

These nice properties come with a price: 
a non-perturbative gauge fixing procedure cannot avoid
Gribov copies~\cite{Gribov_copies},  
and the RI/MOM renormalization conditions may introduce
O($a$) effects into the otherwise on-shell O($a$) improved theory.
Furthermore, the continuum perturbation theory is done
in infinite space-time volume and in the chiral limit,
whereas the numerical simulations are performed at finite quark masses 
and in a finite volume with periodic boundary condition.
To satisfy the RI/MOM conditions 
chiral extrapolations are required and one ought to check for finite
volume effects. Finally, the window 
condition $\Lambda \ll \mu \ll a^{-1}$ may not always be
satisfied. In particular, problems may be expected with operators
which couple to the pion. In fact, the JLQCD collaboration
did not use the axial density in the quark mass renormalization,
as the  renormalization constant could not be
extrapolated to the chiral limit~\cite{MOM5}. This behaviour is
attributed to a contribution of the Goldstone pole which 
should vanish at high enough scales~\cite{MOM}. 
For a more detailed discussion of this problem the reader is referred to 
refs.~\cite{Cudell,MOM4}.

\subsection{Modifications of the RI/MOM scheme}

At this conference, Y.~Zhestkov (RBC collab.) has presented a first
attempt to reach higher scales $\mu$ 
before matching to perturbation theory~\cite{MOM7}.
First results look promising, although it is not
obvious how finite volume, renormalization
and cut-off effects can be disentangled systematically.
The situation could be simplified if the RI/MOM condition
was imposed in a finite volume at constant $\mu L$.
This would define a finite volume scheme, and
finite-size scaling techniques could then be applied to
reach higher scales. However, this means that all the
perturbative calculations would have to be re-done
in the finite volume set-up.

Finally, attempts are being made to implement
off-shell O($a$) improvement of the quark vertex functions
and propagators involved in the RI/MOM scheme.
A short discussion is deferred to sect.~6.

\section{FINITE VOLUME SCHEMES}

\subsection{Basic requirements}

Finite volume schemes are obtained by using the finite
extent of space-time to set the renormalization scale~\cite{LWW,letter},
i.e.~one identifies $\mu=L^{-1}$ in eq.~(\ref{scales}).
Using recursive finite size scaling methods one may 
bridge large scale difference, so that eq.~(\ref{scales})
reduces to $a\ll L$ for any single lattice, and
applicability of perturbation theory, 
$\Lambda \ll \mu$, is required only for the
large scales (i.e.~small $L$) covered by the recursion.

Finite volume schemes can be defined in many ways. 
However, one would like to maintain gauge invariance
and respect on-shell improvement. Furthermore, 
the renormalization conditions
should be easy to evaluate both by numerical simulation 
and in perturbation theory.
A large family of renormalization schemes 
with all of these properties derives
from the Schr\"odinger functional (SF)~\cite{LNWW,StefanI}
and is referred to as SF scheme. 

\subsection{The Schr\"odinger functional}

The SF is the functional integral for 
QCD on a hyper-cylinder, i.e.~all fields are $L$-periodic
in the spatial directions, and satisfy Dirichlet boundary conditions
at Euclidean times $0$ and $T$. For the quark fields one sets
\begin{eqnarray}
  {P_+\psi|}_{x_0=0}=&\!\!\rho,
  \qquad {P_-\psi|}_{x_0=T}=&\!\!\rho',\nonumber\\
  {\bar\psi P_-|}_{x_0=0}=&\!\!\bar\rho,
  \qquad {\bar\psi P_+|}_{x_0=T}=&\!\!\bar\rho'\,,
\end{eqnarray}
with $P_\pm=\frac12(1\pm\gamma_0)$, and the spatial components
of the gauge potential take prescribed values $C_k^{}$ and $C'_k$.
Renormalizability of QCD in this situation has been discussed 
in~\cite{LNWW,StefanII}, with the result 
that the SF is finite after the usual renormalization of the
parameters and multiplicative renormalization of all fermionic boundary fields
by the same renormalization constant.

The SF is a gauge invariant  functional of the boundary fields,
${\cal Z}={\cal Z}[\rho',\rhobar',C';\rho,\rhobar,C]$, 
and correlation functions may be defined as usual
\begin{equation}
  \langle O\rangle = \left\{{\cal Z}^{-1}
   \int_{\rm fields} O\, \rme^{-S}\right\}_{\rho=\rho'=\rhobar=\rhobar'=0},
\end{equation}
where O may contain composite operators and boundary fields 
$\zeta,\zetabar$, which are obtained by differentiating
with respect to the fermionic boundary values, e.g.
$
   \zeta(\bfx) \rightarrow \delta/\delta \rhobar(\bfx).
$
An important point to notice is that the gauge field
boundary condition imply that only global gauge transformations
are allowed at the boundaries. Bilinear 
boundary sources like $\zetabar(\bfx)\Gamma\zeta(\bfy)$ 
are therefore gauge invariant even for $\bfx\ne\bfy$. 
%
\begin{figure}[htb]
\vskip -10ex
\epsfig{file=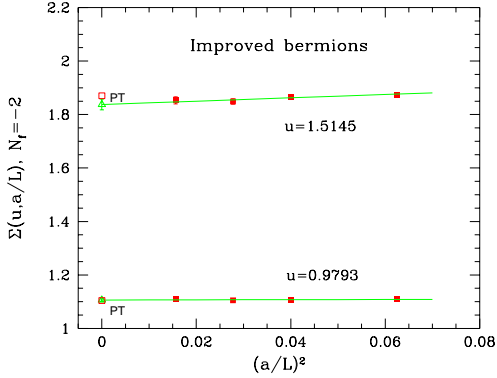,
       width=7cm}%
\vskip -5ex
\caption{Continuum extrapolation of the
step scaling function for the SF coupling with $\Nf=-2$,
using O($a$) improved Wilson ``bermions''~\cite{juri}.}
\label{bermion}
\vskip -4ex
\end{figure}
%
%
\subsection{Renormalized coupling}

A non-perturbatively defined, gauge invariant 
running coupling can be obtained by choosing a 
family of boundary gauge fields depending on a parameter 
$\eta$ and setting~\cite{SU3}
\begin{equation}
  \bar{g}^2(L) = 
  \left.{\partial_\eta\Gamma_0}
  \over{\partial_\eta\Gamma}\right\vert_{\eta=0,m=0}.
\label{SFcoupling}
\end{equation}
Here, $\Gamma(\eta)=-\ln {\cal Z}[0,0,C'(\eta);0,0,C(\eta)]$, 
is the effective action of the SF with 
the perturbative expansion,
\begin{equation}
\Gamma = g_0^{-2} \Gamma_0 + \Gamma_1 + g_0^2\Gamma_2 +\ldots.
\end{equation}
The numerator in eq.~(\ref{SFcoupling}) thus ensures
the standard normalization of a coupling constant.
In eq.~(\ref{SFcoupling}) $T=L$ is assumed,  
the boundary gauge fields are scaled with $L$
and the quark masses are set to zero. 
As a result the SF coupling is a quark mass independent
coupling which runs with the scale $L$.
Its perturbative relation to the $\MSbar$ coupling 
is known to 2-loop order~(cf.~\cite{Bode} and references therein).

For a completely different definition of a finite 
volume running coupling in the SU(2) pure gauge theory 
we refer to ref.~\cite{alphatp}. 

\subsection{Non-perturbative running}

\begin{figure}[htb]
\vskip -25ex
\epsfig{file=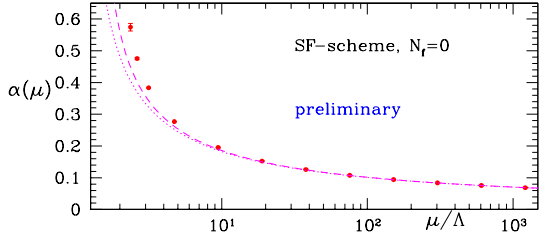,
       width=7.5cm}%
\vskip -10ex
\caption{Non-perturbative running of the SF coupling ($\Nf=0$). For comparison
also the perturbative 2-loop and 3-loop curves are shown.}
\vskip -4ex
\label{alpha_standard}
\end{figure}

The coupling itself is an observable in numerical simulations,
and the theory may now be renormalized 
in the chiral limit by  imposing
\begin{equation}
  \bar{g}^2(L)=u.
 \label{Lzero}
\end{equation}
with some numerical value $u$.
The coupling at scale $2L$ is then also determined and can 
be computed as a function of $u$ as follows.
One chooses a lattice size $L/a$ and tunes $\beta$ such as to
satisfy~(\ref{Lzero}). Then one measures, at the same bare parameters,
the SF coupling on a lattice with size $2L/a$.
This yields a lattice approximant to the desired function,
and by repeating these steps several times with increasing 
resolution $L/a$ one may take the limit,
\begin{equation}
  \bar{g}^2(2L)= \sigma(u)= \lim_{a\rightarrow 0} 
 \Sigma(u,a/L)\bigl\vert_{\bar{g}^2(L)=u,m=0}.
\end{equation}
An example for such an extrapolation 
in QCD with $\Nf=-2$ (improved Wilson ``bermions''~\cite{juri})
is shown in fig.~\ref{bermion}.
Repeating the procedure with different values for $u$ 
the step-scaling function $\sigma(u)$ can be constructed in
some interval $[u_{\rm min},u_{\rm max}]$.
Setting $u_0=u_{\rm min}$, one computes recursively
\begin{equation}
  u_{k}=\sigma(u_{k-1}), \quad k=1,2,\ldots,n.      
\end{equation}
In this way the scale evolution of the coupling is traced non-perturbatively
and can be compared to perturbation theory~(cf. figure~\ref{alpha_standard}).
The ALPHA collaboration has determined the running in quenched 
QCD~\cite{SU3,strange1}
and is currently studying the theories with $\Nf=2$ and 
$\Nf=-2$ quark flavours. The corresponding computer programs for the APE1000 
machine are being optimized, and  the leftmost 3 points
in fig.~\ref{alpha_standard} have in fact been produced by the APE1000 
as a warm-up exercise.
%
 
\subsection{Operator renormalization}

To renormalize a composite operator in the SF scheme
one starts by choosing a non-vanishing correlation
with a boundary source field. In the case of the axial
density $P^a=\psibar\frac{\tau^a}2\gamma_5\psi$ we define
\begin{equation}
  O^a=a^6\sum_{\bfy,\bfz}\zetabar(\bfy)\gamma_5\frac{\tau^a}2\zeta(\bfz),
\end{equation}
and similarly ${O'}^a$ with the primed fields. Given the correlation
functions
\begin{equation}
   \fp(x_0)=-\frac13\langle P^a(x)O^a\rangle,\quad 
        f_1=-\frac1{3L^6}\langle O^a {O'}^a\rangle,
\end{equation}
one may then scale all dimensionfull parameters with $L$,
and define the renormalization constant through the ratio~\cite{PeterStefanI}
\begin{equation}
  \Zp(g_0,a/L)=c{\sqrt{f_1}\over{\fp(L/2)}},
\end{equation}
at vanishing quark masses. Here the ratio has been chosen
such that the boundary field renormalization drops out,
and the constant $c$ ensures  $\Zp=1$ to lowest order perturbation
theory. One may then define the step scaling function for 
the renormalization constant,
\begin{equation}
   \sigma_{\rm P}(u) = \lim_{a\rightarrow 0} 
  \left. {{\Zp(g_0,2L/a)}\over{\Zp(g_0,L/a)}}\right\vert_{\bar{g}^2(L)=u,m=0}.
\end{equation}
The equations,
\begin{equation}
   \Zp(2L)=\sigma_{\rm P}(u)\Zp(L),\quad u=\bar{g}^2(L)
\end{equation}
are then again solved recursively, using the previous result 
for the step-scaling function of the SF coupling~\cite{strange1}.


\subsection{Universality}

\begin{figure}[htb]
\epsfig{file=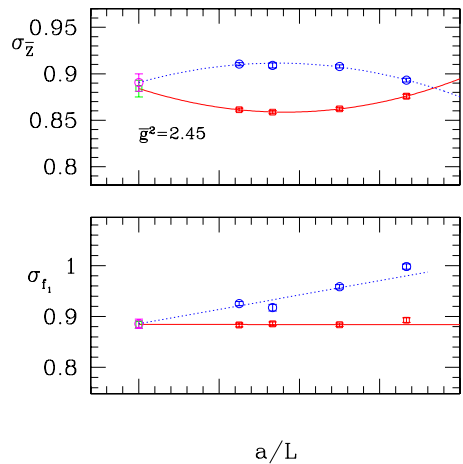,
       width=7cm}%
\vskip -10ex
\caption{The continuum extrapolation of the two step-scaling functions
of ref.~\cite{structure_step} illustrate universality of the continuum limit.
The lattice data were obtained with both O($a$) improved
and unimproved Wilson fermions.}
\vskip -4ex
\label{Karl}
\end{figure}

The step-scaling functions are obtained in the continuum limit
and do not depend on the regularization.
This is illustrated in fig.~\ref{Karl}, 
which shows the continuum extrapolation of  
two step scaling functions, defined in complete
analogy with $\sigma_{\rm P}$.
The simulations  have been performed both with improved and unimproved
Wilson fermions. Due to incomplete improvement residual O($a$) effects are
expected in both cases, and the continuum extrapolated
results nicely agree.~\cite{structure_step}.

In~\cite{Hartmut} an attempt is being made to 
relate the bare chiral condensate 
obtained with Neuberger fermions in~\cite{Pilar} 
to the renormalized axial density in the SF scheme. 
If successful, the step scaling function $\sigma_{\rm P}$
could be used to extract the renormalization 
group invariant chiral condensate.
%
%
\subsection{The static-light axial current}

An interesting new development is the application
of the SF techniques to  QCD with an infinitely
heavy ``static'' quark~\cite{Sommer_Kurth}. This is motivated by the
fact that the $b$-quark is too heavy to be treated
as a relativistic particle whilst maintaining a
physically large volume. Applications to physics
involving the b-quark therefore rely on 
large extrapolations in the quark mass, or on the
use of effective theories, such as non-relativistic
QCD or heavy quark effective theory.
Controlling QCD in the limit of an infinitely heavy 
$b$-quark could lead to a major improvement for quantities
like $F_B$, as the extrapolation in the quark mass is promoted 
to an interpolation.

In ref.~\cite{Sommer_Kurth} the SF for the static light theory 
has been defined in close analogy to the relativistic case.
In particular it was found that the boundary quark fields
are again multiplicatively renormalized. 
The application of Symanzik's improvement programme reveals
that the static action is already O($a$) improved 
and determines the structure of the improved
static light axial current. In the effective 
theory the renormalization of the axial current
is scale dependent and may be performed 
by imposing a similar condition as discussed above for the axial density.
A potential problem is caused by the linear divergence stemming 
from self-energy corrections to the static quark. 
Fortunately, the contribution of this counterterm 
to the static quark propagator is exactly known.
It is then possible to choose the ratios of correlation functions
and parameters such that both the boundary field normalization and 
the self-energy counterterms cancel exactly.

\begin{figure}[htb]
\vskip -10ex
\epsfig{file=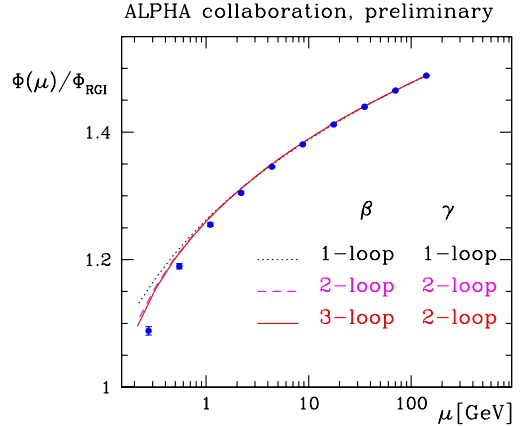,
        clip=,
        angle=0,
       width=8.5cm}%
\vskip -12ex
\caption{Non-perturbative running of the static-light axial current
for $\Nf=0$. For comparison also 
the perturbative curves are shown~\cite{Sommer_Kurth}.}
\vskip -4ex
\label{current}
\end{figure}

The ALPHA collaboration
has obtained  the step-scaling function in the continuum
limit, and the non-perturbative 
running of (a matrix element) of the current
is shown in figure~\ref{current}. The renormalization problem 
has thus been solved and work concerning 
the matrix element $F_B^{\rm stat}$ is in progress.
%
%
%
\section{FINITE RENORMALIZATIONS}

In cases where a non-anomalous continuum symmetry is broken by the
regularization, finite (i.e.~scale independent)
renormalizations are necessary to restore the symmetry up to cutoff effects.
Well-known examples are the space-time symmetry,
which is broken up to the discrete symmetry group of the
space-time lattice, or global axial symmetries, which are
completely broken in lattice QCD with Wilson quarks.
In the case of chiral symmetry, the standard solution
consists in imposing the validity of continuum chiral Ward identities
in the theory at finite lattice spacing~\cite{Bochicchio_Et_Al}.
The same idea has been put foward in the case of supersymmetry
by Curci and Veneziano~\cite{Curci_Veneziano}, and first numerical
and perturbative results along these lines have been presented
at this conference~\cite{Feo,Farchioni}.
A similar treatment of the O(4) Euclidean space-time symmetry
appears difficult in practice, as it requires the
construction of the energy momentum tensor,
which is a non-trivial problem in itself~\cite{Menotti_et_al}.
I will focus on the more familiar problem
of explicit chiral symmetry breaking and its consequences.

\subsection{Chiral symmetry and Wilson fermions}

The  explicit breaking of chiral symmetry by the Wilson term
entails additive quark mass renormalization, 
a non-trivial renormalization of the non-singlet axial current,
\begin{equation}
  (A_{\rm R})^a_\mu = \Za(g_0) A_\mu^a,
\end{equation}
and mixing of operators with the wrong nominal chirality.
For example, the $\Delta S=2$ operator
\begin{equation}
   O^{\Delta S=2} = \sum_{\mu}
                    \left(\bar{s}\gamma_\mu(1-\gamma_5)d\right)^2,
\end{equation}
decomposes in parity even and parity odd parts,
\begin{equation}
  O^{\Delta S=2} = O_{\rm VV+AA} - 2\,O_{\rm VA}.
\end{equation}
which are renormalized differently.
While the parity odd operator is multiplicatively 
renormalized~\cite{Bernard_et_al},
the parity even operator mixes with four other operators
of the same dimension, 
\begin{equation}
   O_{\rm VV+AA}^{\rm sub}= O_{\rm VV+AA} +\sum_{i=1}^4 z_i(g_0)O_i^{[d=6]}, 
\label{weakmix}
\end{equation}
and the subtracted operator is then multiplicatively renormalized.

In order to determine the scale independent
renormalization constants $\Za$,$z_i$,$Z_{VV+AA}/Z_{VA}$, etc.,
Bochicchio et al.~\cite{Bochicchio_Et_Al} 
have suggested to impose continuum axial Ward identities (AWI)
as normalization conditions. A generic continuum 
AWI in the theory with $\Nf$ mass degenerate quarks
has the form
\begin{eqnarray}
 \int_R\rmd^4x \left\langle
        \left(\partial_\mu A_\mu^a(x)-2mP^a(x)\right)O(y)O_{\rm ext}
               \right\rangle \nonumber\\ 
  +\left\langle\Bigl(\delta_{\rm A}^aO(y)\Bigr)O_{\rm ext}
   \right\rangle = 0, 
 \label{AWI}
\end{eqnarray}
where $R$ is a finite space time region containing $y$, and
$O_{\rm ext}$ is an arbitrary product of fields localized outside $R$.
A simple example is the PCAC relation which is obtained for
$O(y)=\delta(y-x)$. If imposed on the lattice
it  determines the additive quark mass renormalization constant.
Considering  axial variations of the axial current,
the AWI's determine the normalization constant $\Za$. 
More generally the axial Ward identities determine
the relative renormalization of operators belonging to the same
chiral multiplet, e.g.~the scalar and axial densities, 
or $O_{\rm VA}$ and $O_{\rm VV+AA}$.

Applying AWI's has become standard in the quenched approximation.
At this conference, R.~Horsley has presented
a first attempt to  determine  $\Za$ 
for $\Nf=2$ flavours of O($a$) improved Wilson fermions~\cite{Horsley}.

The AWI's can also be implemented more indirectly by applying 
the RI/MOM scheme to finite renormalization
constants. With the appropriate choice
for the renormalization conditions the AWI's
are automatically satisfied at large enough renormalization 
scales~\cite{RIMOMfinite}. 

\subsection{Systematic errors and O($a$) ambiguities}

There is an infinity of axial Ward identities, and the results
for a given scale independent renormalization constant
will differ at O($a$) or O($a^2$), if the theory is O($a$) improved.
One may be tempted to assign a systematic error to the renormalization 
constant corresponding to a ``typical spread'' of the results.
This is somewhat subjective and unsatisfactory, in particular when
the spread is large. 

An alternative point of view has been advocated
in refs.~\cite{paperIV,babp}. One {\em defines}
a given finite renormalization constant $Z$ by imposing the Ward 
identity at fixed renormalized parameters. 
As a result one obtains a smooth function of the bare coupling, $Z(g_0)$, 
with relatively small errors. 

When $Z(g_0)$ is used in a continuum extrapolation
the O($a$) ambiguity disappears
in the limit. This can also be checked explicitly
by choosing another renormalization condition which 
yields a different function $\tilde{Z}(g_0)$. The difference
$Z(g_0)-\tilde{Z}(g_0)$ must then vanish
in the continuum limit, with an asymptotic 
rate proportional to $a$ or $a^2$. 
In ref.~\cite{babp} the ALPHA collaboration 
has confirmed this expectation with 
the example of the finite renormalization constant $Z_m\Zp/\Za$. 

\subsection{Avoiding finite renormalizations}

Some of the scale-independent renormalizations with Wilson type quarks
can be avoided by working in a different basis of fields. 
Such an approach has been presented at last year's 
conference~\cite{lat99_tmQCD}, and has been dubbed 
twisted mass QCD (tmQCD).
One considers $\Nf=2$ quark flavours with 
the lattice action,
\begin{equation}
   S_{\rm F} = a^4\sum_x \psibar(D_W+m_0+i\muq\gamma_5\tau^3)\psi.
\end{equation}
Here, $D_W$ is the massless Wilson-Dirac operator,
$m_0$ is the standard bare mass, and $\muq$ is referred to as 
twisted mass parameter. This action has already appeared 
in the literature in different contexts~\cite{Aoki_tmQCD,BardeenII}.
The new idea consists in renormalizing the theory at finite 
twisted mass, and in a re-interpretation of the renormalized
theory as standard QCD with $\Nf=2$ mass degenerate quarks.
If the renormalization scheme is chosen with care,
the formulae relating renormalized tmQCD and standard QCD 
can be inferred from classical continuum 
considerations~\cite{work_in_progress}.

We thus  consider the classical continuum lagrangian
\begin{equation}
    {\cal L}_f=\psibar\left(D\kern-7pt\slash
                +m+i\mu_q\gamma_5\tau^3\right)\psi 
               + \bar{s}\left(D\kern-7pt\slash +m_{\rm s}\right)s.
\end{equation}
where $\psi$ is a doublet of light quarks, 
and we have also included the strange quark.
A  chiral (non-singlet) rotation of the doublet fields,
\begin{eqnarray}
 \psi'    &=&\exp(i \alpha\gamma_5\tau^3/2)\psi,\nonumber\\
 \psibar' &=&\psibar\exp(i \alpha\gamma_5\tau^3/2),\label{axialI}
\end{eqnarray}
with  $\tan\alpha = \muq/\mq$ transforms the 
Lagrangian to its standard form,
\begin{equation}
  {\cal L}'_f(x) =\psibar'(x)\left(D\kern-7pt\slash
                +m'\right)\psi'(x) +  
          \bar{s}\left(D\kern-7pt\slash +m_{\rm s}\right)s, 
\end{equation}
with the light quark mass $m'= (m^2+\muq^2)^{1/2}\,$.
The rotation of the quark and anti-quark fields 
also induces a rotation of the composite fields.
For quark bilinears containing only the light quarks
one finds e.g.
\begin{eqnarray}
  {A'}_\mu^1 &=& \cos(\alpha) A_\mu^1 + \sin(\alpha) V_\mu^2,\\  
  (\psibar\psi)' &=& \cos(\alpha) \psibar\psi + 2i\sin(\alpha)P^3,
\end{eqnarray}
and the $\Delta S=2$ operator behaves as follows,
\begin{equation}
  O_{\rm VV+AA}' = \cos(\alpha)\,O_{\rm VV+AA} 
 -2 i\sin(\alpha)\, O_{\rm VA}.   
\end{equation}
The fields in the primed basis refer to the standard QCD basis, 
and the primed composite operators are hence interpreted as
usual. Working in the twisted basis at $\alpha=\pi/2$
one may use the vector current
to compute $F_\pi$, obtain the chiral condensate 
from $P^3$ and compute the $K-\bar K$ mixing
amplitude using the multiplicatively renormalizable 
operator $O_{\rm VA}$~\cite{Bernard_et_al}. 
Work on tmQCD by the ALPHA collaboration
is in progress, and the results of a scaling
test have been presented at this conference~\cite{Michele}.

A variant of the above proposal concerning $K-\bar K$ mixing 
can also be realized with standard Wilson 
fermions~\cite{Martinelli_BK}, by using  
the axial Ward identity~(\ref{AWI}). One
considers an axial variation of $O=O_{\rm VA}$,
\begin{equation}
  \delta_{\rm A}^3 O_{\rm VA} = -\frac12 O_{\rm VV+AA},
\end{equation}
and chooses $O_{\rm ext}$ as the product of interpolating fields 
for $K$ and $\bar K$. Imposing the AWI on the lattice 
is equivalent to defining the correlation function involving 
$O_{\rm VV+AA}$ through the remainder
of the AWI, which contains the operator $O_{\rm VA}$.
The trick here is that one directly obtains the desired
matrix element by choosing $O_{\rm ext}$ appropriately.
As in the tmQCD proposal this avoids
solving the complicated mixing problem~(\ref{weakmix}),
and one is left with a scale dependent multiplicative renormalization
for $O_{\rm VA}$. First feasibility tests have been performed and 
presented at this conference~\cite{Martinelli_talk}.

\subsection{Ginsparg-Wilson and Domain-Wall quarks}

Massless Ginsparg Wilson fermions are implicitly characterized by a
Dirac operator which satisfies 
the Ginsparg-Wilson relation~\cite{GW}
\begin{equation}
  D \gamma_5+\gamma_5 D=a D\gamma_5 D.
\label{GW}
\end{equation}
The Ginsparg Wilson (GW) relation implies an exact chiral 
symmetry~\cite{LuscherI}, and hence none of the 
finite renormalization constants discussed in this section is 
needed~\cite{Hasenfratz_mix}.
%
\begin{figure}[htb]
\epsfig{file=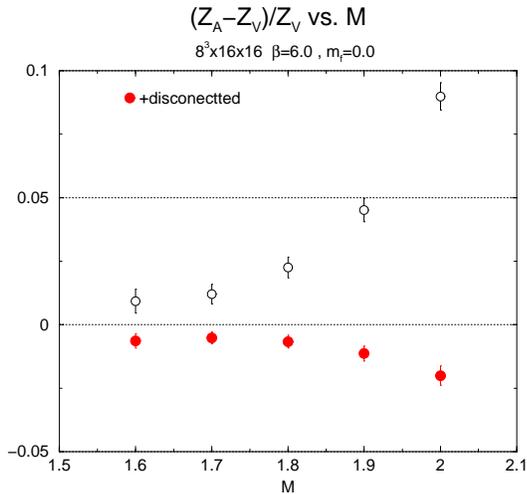,
       width=7cm}%
\vskip -5ex
\caption{Difference between the axial and the vector current normalization
constants as a measure of chiral symmetry violation~\cite{Aoki_DWF}.}
\vskip -4ex
\label{Aoki_ZVZA}
\end{figure}
%

An explicit solution of the Ginsparg-Wilson relation has
been given by Neuberger~\cite{Neuberger}.
Unfortunately, it is computationally very demanding to implement 
GW quarks exactly, and the question arises whether an
approximation may be good enough in practice. A  popular 
choice are Domain Wall fermions~\cite{DWF}, which are
formulated in five dimensions. The Dirac operator 
of the corresponding four-dimensional effective theory 
becomes an exact solution of the  Ginsparg-Wilson relation
as the number of lattice sites in the fifth dimension tends
to infinity. As the approach is expected to be
exponential, one may hope that a small number $N_s$ of 
points in the fifth dimension may already be sufficient.

To check the size of chiral symmetry violations one
may pursue the same strategy as in the Wilson case, 
i.e.~one measures to what extent chiral continuum Ward identities
are violated. 
For example, one may measure the difference between
vector current and axial current renormalization constants
The CP-PACS collaboration has measured both
renormalization constants using Ward identities 
at $m_f=0$, on a lattice of size 
$8^3\times 16\times 16$~(cf. fig.~\ref{Aoki_ZVZA}). 
The authors use the methods of ref.~\cite{paperIV},
and therefore impose SF boundary conditions in the physical
time direction. Note, however, it that
it is not obvious whether this is equivalent
to imposing SF boundary conditions in the effective
4-dimensional theory. 
%
\begin{figure}[htb]
\vskip -15ex
\epsfig{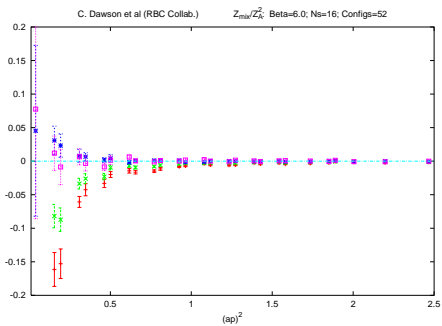}%
\vskip -10ex
\caption{The wrong chirality mixing coefficients for the 
$\Delta S=2$ operator with domain wall fermions, as determined 
with the RI/MOM scheme~\cite{MOM8}.}
\vskip -4ex
\label{Dawson}
\end{figure}
%
The Riken-Brookhaven-Columbia (RBC) collaboration has studied 
the mixing properties of the operator $O_{\rm VV+AA}$~\cite{MOM8}.
They follow ref.~\cite{APE_BK}
and determine the wrong chirality mixing coefficients in a RI/MOM scheme.
The result for $N_s=16$ is shown in fig.~\ref{Dawson}, and one 
observes that the coefficients are indeed small, 
i.e.~a factor of 5-10 smaller
than the corresponding result with (unimproved) Wilson quarks~\cite{Aoki_BK}.

The overall impression is that chiral symmetry violations
seem to be  numerically small in many cases.
However, it is also possible to determine the exponent of the
exponential approach to a Ginsparg-Wilson solution
directly by solving a generalized eigenvalue 
problem~\cite{Heller_Edwards,Hernandez}.
It turns out that the exponent may be rather small at typical
values of $\beta$, and with Wilson's plaquette action 
for the gauge fields. This is potentially dangerous,
and some modification of the original formulation 
of Domain Wall Fermions may be necessary
(cf.~\cite{Vranas} for further details).

\section{O($a$) IMPROVEMENT}

The purpose of Symanzik improvement~\cite{SymanzikI,SymanzikII,OnShell} 
is to accelerate the continuum approach of renormalized quantities
by introducing appropriate counterterms to both the action and
the composite operators of interest.
The counterterm structure follows from the
symmetries of the regularized theory. Without loss of physical 
information improvement may be restricted to 
on-shell quantities, i.e.~particle masses, energies and
correlation functions of local operators which keep
a physical distance from each other. In this case the
equations of motion may be used to reduce the 
number of counterterms.

\subsection{On-shell O($a$) improvement with Wilson quarks}

The on-shell O($a$) improved action for $\Nf$ mass degenerate
Wilson quarks is given by~\cite{SW},
\begin{equation}
  S_{\rm F} = a^4\sum_x\psibar (D_W+m_0 
  +\csw a \frac{i}4\sigma_{\mu\nu}F_{\mu\nu}) \psi,
\end{equation}
where $D_W$ denotes the standard Wilson-Dirac operator.
Other counterterms can be absorbed in the redefinition
of the bare parameters. With the subtracted bare 
quark mass $\mq=m_0-\mc$, these are~\cite{paperI}, 
\begin{eqnarray}
   \gtilde^2  &=& g_0^2(1+\bg a\mq), \\
  \mqtilde   &=& \mq(1+ b_{\rm m} a\mq),  
\end{eqnarray}
and renormalized O($a$) improved parameters in a mass independent
renormalization scheme take the form
\begin{eqnarray}
   g_{\rm R}^2 &=&  \gtilde^2\Zg(\gtilde^2,a\mu),\\
   m_{\rm R}   &=&  \mqtilde\Zm(\gtilde^2,a\mu).
\label{m_renorm}
\end{eqnarray}
Improved composite operators are treated similarly. We restrict attention
to non-singlet quark bilinear operators, e.g.~the axial
current,
\begin{equation}
 (A_{\rm R})_\mu^a = \Za(1+\ba a\mq)\left\{A_\mu^a
   +\ca a\partial_\mu P^a\right\}.
\end{equation}
The vector current and the tensor density have similar additive
counterterms, and fields are multiplicatively rescaled with
a quark mass dependent term and the appropriate $b$-coefficient.
Note that all renormalization constants are functions of the
improved bare coupling $\gtilde$, whereas the improvement coefficients
depend on $g_0$.

\subsection{O($a$) improvement and chiral symmetry}

It is an important observation that all of the above counterterms 
conflict with  chiral symmetry, and thus repair cutoff effects
which originate from  explicit chiral symmetry breaking.
For this reason, lattice QCD with Ginsparg-Wilson 
quarks is on-shell O($a$) improved,
provided the mass term is introduced in the correct 
way~\cite{Niedermayer_98}.

Concerning  Wilson quarks it is a natural 
question whether axial Ward identities can
be imposed as an improvement condition
to determine the coefficients non-perturbatively.
For those coefficients which already appear in the chiral
limit (the $c$-coefficients), this is indeed the case:  the
integral over the region $R$ in eq.~(\ref{AWI})
can be converted into a surface integral over $\partial R$, and
the axial Ward identity becomes an identity 
involving on-shell correlation functions only. 
Away from the chiral limit the answer is more subtle.
The Ward identity now contains an off-shell correlation
function, which is not expected to be improved.

An exception is the PCAC relation, which 
leads to an alternative definition of 
a renormalized O($a$) improved quark mass~(\ref{m_renorm}).
Requiring equality of the two mass definitions then determines
the combination
\begin{equation}
  \bm-\ba+\bp +  g_0^2\, {\partial\ln Z \over\partial g_0^2}\,\bg,
\end{equation}
where $Z=\Zm\Zp/\Za$ is a finite renormalization constant.
Another special case is the coefficient $\bv$ which
may be determined from the vector Ward identities. Alternatively
one may define an O($a$) improved current starting
from the conserved current of the exact flavour symmetry, for
which one finds $\Zv=1$ and $\bv=0$.

It is natural to ask whether the situation can be improved
by allowing for mass non-degenerate quarks.
Unfortunately the general counterterm structure becomes
very complicated, i.e.~there are many more
possible $b$-coefficients~\cite{Sharpe99}. 
Nevertheless, in the case $\Nf\ge 3$ these authors conclude that, 
assuming that $\bg$ is known, all but three 
combinations of $b$-coefficients are indeed determined by
Ward identities.

An independent proposal to determine the $b$-coefficients
is based on the idea that chiral
symmetry is restored at short distances~\cite{Talevi}. 
This implies that the physical quark mass dependence 
is invariant under a sign change of the mass. 
As the $a\mq$ cutoff effects violate this invariance, one
may in principle be able to isolate the coefficients. 
Unfortunately, in order to resolve sufficiently short distances 
one may need very small lattice spacings so that the method
may not apply in the interesting range of $\beta$-values.

The situation becomes much simpler if one restricts
attention to the quenched approximation. The coefficient
$\bg$ then vanishes and an improved bare quark mass parameter
$\mqtilde=\mq(1+\bm a \mq)$ is defined for each flavour 
individually. Furthermore, the flavour off-diagonal 
quark bilinear fields are improved with the same $b$-coefficients as in the
degenerate case, but with $\mq$  replaced by the 
average of the valence quark masses in 
the operator~\cite{GiuliaRoberto,Gupta1}. As shown in ref.~\cite{Gupta1}
it is then indeed possible to determine all
$b$-coefficients from axial Ward identities.
In particular, off-shell correlation functions in the axial Ward identity
are avoided by considering a partial chiral limit with two massless
flavours and a further quark mass which is accompanied by 
the $b$-coefficient of interest.

\subsection{Results}

\begin{figure}[htb]
\epsfig{file=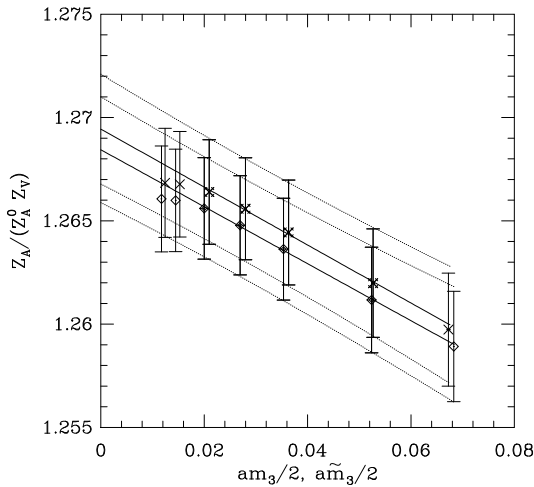,
       width=7cm}%
\vskip -5ex
\caption{The combination $\ba-\bv$ at $\beta=6.2$ 
is determined by the slope~\cite{Gupta_opus}.}
\vskip -5ex
\label{gupta1}
\end{figure}

In perturbation theory the improvement coefficients
are known to one-loop order~\cite{Wohlert,paperII,StefanRainer,PeterStefan}.
To this order the coefficients do not depend on $\Nf$ except
for $\bg$~\cite{StefanRainer}.
Nonperturbative results have been obtained mostly
in the quenched approximation with the exception of 
$\csw(\Nf=2)$~\cite{Karl_Rainer}.
Numerical results exist for $\csw$ ~\cite{paperIII,Klassen},
$\ca,\bv$~\cite{paperIV} and $\cv$~\cite{MarcoRainer}. 
All these results were obtained by exploiting chiral 
Ward identities formulated in the Schr\"odinger functional,
and in general the results are obtained as functions of $g_0$,
for $\beta \ge 6.0$.
In ref.~\cite{GiuliaRoberto} the combinations $\ba-\bp$
and $\bm$ were determined for values $\beta\ge 6.2$,
by considering the PCAC relation with non-degenerate quarks.
Recently Bhattacharya et al.~\cite{Gupta_opus} have
published the final version of their work with numerical results for
all improvement coefficients needed for the improvement of
quark bilinear operators. The numerical results are obtained 
at two $\beta$ values, $6.0$ and $6.2$, using hadronic
correlation functions. An example is given in fig.~\ref{gupta1}.
%
%
\begin{figure}[htb]
\epsfig{file=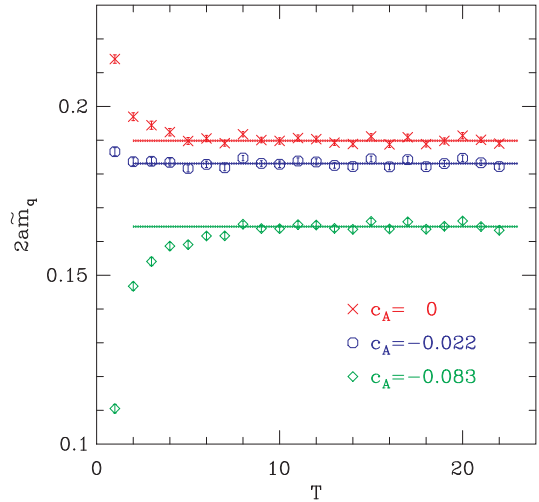,
       width=7cm}%
\vskip -5ex
\caption{The improvement condition for the coefficient $\ca$ 
at $\beta=6.0$~\cite{Gupta_opus}.}
\vskip -4ex
\label{gupta2}
\end{figure}
%
%
\begin{figure}[htb]
\vskip -10ex
\epsfig{file=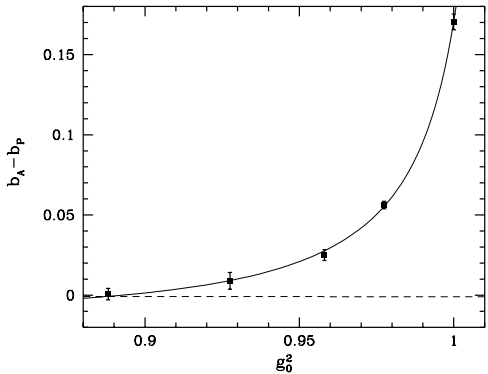,
       width=7cm}%
\vskip -5ex
\caption{Simulation results and fit function for $\ba-\bp$, determined
from an improvement condition imposed at constant physics~\cite{babp}.}
\vskip -4ex
\label{babp}
\end{figure}

\begin{figure}[htb]
\vskip -10ex
\epsfig{file=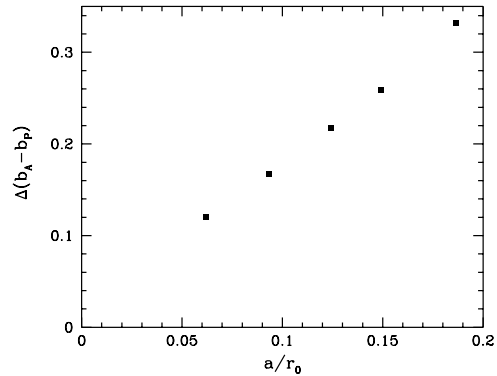,
       width=7cm}%
\vskip -5ex
\caption{The difference between two different determinations 
of $\ba-\bp$ at constant physics is plotted
against $a/r_0$. One observes the expected decrease as the continuum 
limit is approached~\cite{babp}.}
\vskip -4ex
\label{deltababp}
\end{figure}

In general, the agreement with previous results 
is reasonably good, except for $\cv$ and $\ca$ at $\beta=6.0$. 
In particular, their preferred value
for $\ca$ is about $-0.04$ which is roughly half of the value 
obtained in~\cite{paperIII}. Fig.~\ref{gupta2} illustrates the sensitivity
of the method: It shows the PCAC mass obtained from 2-point
functions as a function of time separation. 
The excited states in the pion channel are sensitive
to $\ca$ which is tuned such as to extend the plateaux to smaller times. 
This procedure is repeated for 
various quark masses and an extrapolation to the chiral limit is performed.
Similar results for $\ca$ with essentially 
the same method have been obtained in ref.~\cite{Collins_Davies}.

Improvement coefficients are intrinsically ambiguous by terms
of O($a$). While a large O($a$) ambiguity in the case of $\ca$ cannot
be excluded, it may also be that the chosen improvement condition
is afflicted by large higher order lattice artifacts,
resulting in an artificial enhancement of the O($a$) ambiguity.
In ref.~\cite{paperIII} some checks were carried out to ensure
that this does not happen and some alternative 
improvement conditions have corroborated
the previous result for $\ca$ at $\beta=6.0$~\cite{ALPHA_ca}. 

The ALPHA collaboration has extended the work of~\cite{GiuliaRoberto}
and determined $\bm$ and $\ba-\bp$ covering a range of values
$\beta \ge 6.0$. Surprisingly, the $\ba-\bp$ coefficient
shows a rather large O($a$) ambiguity. It was
therefore decided to define the improvement condition
at ``constant physics'', keeping all quark masses and
scales fixed in units of a physical scale (cf.~sect.~5).
The result for the final choice of the improvement condition
is shown in figure~\ref{babp}.  
As expected the difference to an alternative definition
decreases with a rate roughly $\propto a$~(cf.~fig.~\ref{deltababp}).

\subsection{Off-shell improvement of the gauge fixed theory}

In order to render the RI/MOM scheme compatible
with O($a$) improvement, one needs to extend
improvement  to the gauge fixed theory (Landau gauge), 
and to off-shell quantities, such as the vertex functions
and quark propagators  entering the RI/MOM 
renormalization conditions. 

According to the authors of ref.~\cite{Rossi_preliminary}
(cf. also~\cite{Lubicz_prop}), 
this may be achieved by using the {\em on-shell} 
improved action supplemented with a gauge fixing 
term and the action of the ghost fields. Furthermore, one
introduces  additional O($a$) counterterms
to the composite operators and the quark and anti-quark fields.
In the case of gauge invariant quark bilinear operators
the only additional counterterms are the ones which vanish by the
equations of motion. Improvement of the quark field also 
requires the introduction of a non-gauge invariant counterterm $\propto a 
\partial{\kern -5pt}\slash\psi$ and an analogous term for 
the anti-quark field.

The latter counterterm is not needed in explicit
one-loop calculations carried out in ref.~\cite{Schierholz_offshell}.
These authors organize the calculation differently, by
explicitly subtracting contact terms in the quark propagator
and the vertex functions. However, this seems to be equivalent
to the effect of those counterterms which vanish by the equations
of motion. Finally we mention that some of the
additional improvement coefficients for quark bilinear operators
have been determined non-perturbatively in ref.~\cite{Gupta_opus}.

\section{CONCLUDING REMARKS}

In the last few years a whole arsenal of tools has
been created which will help to ultimately 
solve the non-perturbative renormalization problem
of QCD and similar theories. In this talk I have tried
to give an impression of the problems and the conceptual
and technical progress in the field. 

An important omission in this review concerns power divergences.
The reason is that power divergences do not 
only present a technical but above all a serious conceptual challenge.
To illustrate the potential problem, 
assume that QCD is regularized on the
lattice with Ginsparg-Wilson quarks,  and suppose we want
to renormalize the isosinglet scalar density at non-zero
quark mass. The structure of the renormalized operator
is determined by the symmetries,
\begin{equation}
  S^0_{\rm R} = Z_{\rm S}
  \left\{S^0 + m a^{-2} c_{\rm S}(am,g_0)\right\},  
\end{equation}
where $c_{\rm S}$ is a dimensionless coefficient which depends on the
bare parameters and is independent of the renormalization 
scale $\mu$~\cite{Testa_broken}.
To define the renormalized operator one should be able to 
impose a renormalization condition which does not refer to 
the regularization. In the chiral limit this is 
not a problem, as the additive renormalization vanishes exactly.
At non-zero mass, however, one must first define a renormalized
quark mass, which is only possible up to an intrinsic O($a^2$)
ambiguity. The renormalization condition for the scalar density
now refers to the renormalized quark mass, and one
may be worried  that the O($a^2$) ambiguity combines
with the quadratic divergence to produce an O(1) ambiguity
in the ``renormalized operator''.

The example of the isosinglet scalar density may seem
academic, but  similar  problems are expected in the renormalization
of the effective weak hamiltonian. A theoretical solution to this
problem might be Symanzik improvement to higher orders. However,
this requires the introduction of four-quark operators 
in the action and does not seem very practical.
New ideas may be necessary to either solve or circumvent
these difficult renormalization problems,
and an interesting new approach has already been 
proposed~\cite{Martinelli_OPE}.

\vskip 1ex

I would like to thank the conference  organizers  for the opportunity
to give this talk. I have benefitted from discussions with many
participants at the conference, and in particular with
M. L\"uscher,  R. Petronzio, G.C. Rossi and R. Sommer. 
Furthermore I thank R. Sommer for critical comments on 
a draft of this writeup.
Support by the European Commission under
grant No.~FMBICT972442 and by CERN is gratefully acknowledged.

\end{document}